\documentstyle[11pt,epsfig,amssymb]{article}
\newlength{\dinwidth}
\newlength{\dinmargin}
\newcommand {\nn} {\nonumber}
\newcommand {\half} {\frac{1}{2}}
\newcommand {\p} {\prime}
\newcommand {\G} {{\cal G}}

\newcommand{\R}{\bar{R}}

\newcommand{\T}{{\cal T}}

\newcommand{\be}{\begin{equation}}
\newcommand{\ee}{\end{equation}}
\newcommand{\bea}{\begin{eqnarray}}
\newcommand{\eea}{\end{eqnarray}}

\renewcommand{\theequation}{\arabic{equation}}

\begin{document}

\thispagestyle{empty}
\addtocounter{page}{-1}
\vspace*{2.0cm}

\centerline{\Large \bf Warped brane-world compactification}
\vspace*{0.2cm} \centerline{\Large \bf with Gauss-Bonnet term}
\vspace*{0.6cm} \centerline{\bf Y. M. Cho ${}^{a}$ \mbox{and}
Ishwaree P. Neupane ${}^{a, b}$}

\vspace*{0.3cm}

\centerline{\it ${}^a$ School of Physics and Center for
Theoretical Physics} \centerline{\it Seoul National University,
Seoul 151-742 Korea} \vspace*{0.3cm}

\centerline{\it ${}^{b}$ The Abdus Salam ICTP, Strada-Costiera,
11-34014, Trieste, Italy}

\vspace*{0.3cm}

\vspace*{0.8 cm}
\centerline{\bf Abstract} \vspace*{0.4cm}

In the Randall-Sundrum (RS) brane-world model a singular
delta-function source is matched by the second derivative of the
warp factor. So one should take possible curvature corrections in
the effective action of the RS models in a Gauss-Bonnet (GB) form.
We present a linearized treatment of gravity in the RS brane-world
with the Gauss-Bonnet modifications to Einstein gravity. We give
explicit expressions for the Neumann propagator in arbitrary $D$
dimensions and show that a bulk GB term gives, along with a tower
of Kaluza-Klein modes in the bulk, a massless graviton on the
brane, as in the standard RS model. Moreover, a non-trivial GB
coupling can allow a new branch of solutions with finite Planck
scale and no naked bulk singularity, which might be useful to
avoid some of the previously known ``no--go theorems" for RS
brane-world compactifications.

\vspace*{0.6cm}

\begin{flushleft}
\end{flushleft}
\vspace*{0.2cm}
\begin{flushleft}
\end{flushleft}

\vspace*{0.6cm}

\baselineskip=18pt

\newpage

\setcounter{equation}{0}

\section{Introduction}
In recent years, there has been considerable interest in the
dynamics of brane interactions and brane-worlds with warped extra
dimensions. The interest was motivated by the ideas coming out of
the Randall-Sundrum (RS) brane-world scenario with a warped fifth
dimension~\cite{Randall1,Randall2}. The RS brane-world scenario
promptly received many generalization in higher dimensions
~\cite{ArkaniHamed99b,Garriga99a,Kaloper99a,Csaki00a,Giddings00a}
which have subsequently attracted much interest in gravity and
cosmology research. A necessary ingredient of the brane-world
models is that the space-time metric contains a warp factor
$e^{-\,2A(z)}$ that depends non-trivially on the extra
dimension(s), and, as a result, a brane model of two $3$-branes
with opposite brane tensions~\cite{Randall1}, known as RS1 model,
may provide a geometrical resolution of hierarchy problem. This
proposal was made more concrete in the scenario pioneered further
by Randall and Sundrum, known as the RS single brane (or RS2)
model~\cite{Randall2}, where the fifth dimension is non-compact.
The latter model, viewed as an alternative to Kaluza-Klein
compactification~\cite{Randall2}, gives an illustrative example of
localized gravity on a singular $3$-brane.

To localize gravity to the RS brane, one considers a bulk
space-time with negative cosmological constant. The low energy
effective action in five space-time dimensions is taken to be \be
\label{eq1} S=\int_{{\cal B}}
d^5X\,\sqrt{|g|}\left(\frac{R}{\kappa_5}-2\Lambda\right)
+\sum_{i}\int^{i'th~brane}_{\partial{\cal B}} d^4x\, \sqrt{|h|}\,
\left({\cal L}_m-\T_i\right)\,, \ee where $\kappa_5=16\pi
G_{(5)}=M_{(5)}^{-3}$, with $M_{(5)}$ being the five-dimensional
mass (energy) scale, and $h$ is the determinant of the induced
metric on the $3$-brane.  In a single brane setup, the brane
tension $\T$ is positive ($\T>0$), and the anti-de Sitter length
scale $\ell$ is set by a relation $\ell^2=-\,6M_{(5)}^3/\Lambda$.
One of the interesting features of the RS action~(\ref{eq1}) is
the presence of four-dimensional gravity as the zero-mode spectrum
of a five-dimensional theory. One also finds correct momentum and
tensor structures for the graviton propagator due to the
``brane-bending''
mechanism~\cite{Garriga99a,Giddings00a,JEKim00b,IPN01b}.

The action~(\ref{eq1}) describes the dynamics of a background
metric field for sufficiently weak curvatures and sufficiently
long distances. To further explore the general properties of
brane-worlds, it is more natural to consider the leading order
curvature corrections as predicted by string (bulk) theory. With
warped space-time metrics in the bulk, however, one should take
the curvature corrections in the RS action to be no larger than
the second derivatives of the metric. Thus one can introduce the
higher order corrections only in a special Gauss-Bonnet (GB)
combination. The GB term arises as $\alpha'$ corrections in
bosonic string theory~\cite{Deser85a,Metsaev87a} and in heterotic
M-theory scenario of Horova-Witten type~\cite{Kashima}. Such
corrections might be crucial in space-time dimensions $D\geq 5$,
in particular, when the brane-world scenario is viewed as a low
energy limit of string/M theory.

There are now growing interest in the RS brane-world models
modified by higher derivative corrections. Such corrections in a
Gauss-Bonnet form, for constant dilaton fields, had been
considered earlier in Refs.~\cite{JEKim00a,Zee}, see also
Ref.~\cite{Nojiri00c} which give some realizations of brane-world
inflation due to quantum correction. It is learnt that brane-world
configurations with a GB term and several co-dimensions one branes
can induce brane junctions of non-trivial
topology~\cite{JEKim01a,IPN01c}. Furthermore, the presence of a GB
term coupled to a bulk scalar in the effective action leads to
interesting physics in a variety of context, ranging from gravity
localization~\cite{IPN01a,Mavro,IPN00,Deruelle,Collins,
Giovannini,Kaku00a,kaku00b,0106203} to FRW type cosmology on a
brane~\cite{Nojiri00d,Germani02a,IPN02a}. It is worth noticing
that a five-dimensional brane-world model with a GB term
reproduces all essential properties of the original RS models,
including the tensor structure of a massless
graviton~\cite{IPN01b}. In this paper, we extend the work in Ref.
~\cite{IPN01b} to the $D$ dimensional space-times, and also
examine the general properties of brane-world solutions without
and with a bulk scalar.

A fine tuned relation between the bulk cosmological constant and
the brane tension is required in the RS model to maintain flatness
of the $3$-brane~\cite{CEGH}. This problem has been known as
"no-go theorem"~\cite{CEGH,Freedman99a,Malda00a} for non-singular
RS or de Sitter compactifications based in Einstein gravity. It is
therefore of interest to know whether these arguments can be
changed to cover alternatives in the brane-world actions. With a
GB correction to Einstein gravity, fine tuning is required in the
RS models to get a singularity free solution with the finite
Planck scale~\cite{Rama02a}. Nevertheless, in the presence of such
interaction, there exists a new branch of the solutions, for which
it might be possible that the only bulk singularity occurs when
the warp factor $e^{-A(z)}$ vanishes at the anti-de Sitter
horizon, $z\to \infty$. Presumably, a GB term might smooth out the
bulk singularities and hence avoid some of the previously given
no-go arguments for the RS compactifications.

The rest of the paper is organized as follows. In Sec. 2, we
introduce the brane-world action and find intersecting brane
backgrounds with more than one uncompactified extra dimensions. We
give the basic expressions of the linearized Einstein equations
with a warped metric, and generalize results of the RS2 model when
there are two transverse directions. We then discuss in Sec. 3
some interesting features of the RS solutions modified by a GB
term. In Sec. 4, we find the Neumann propagators in $D$
dimensions, where the background is described as a $3$-brane
embedded in AdS space of dimensions $D\geq 5$. In Sec. 5 we
analyze the energy conditions and compare the RS solutions in five
dimensions with those arising due to a non-trivial GB coupling. In
Sec. 6 we give some insights on the nature of singularities or
no-go theorem for a class of brane-world gravity models coupled to
a bulk scalar field and a Gauss-Bonnet self-interaction term.
Section 7 contains discussion and outlooks. In the Appendices, we
give some useful derivations of the linearized equations for a
class of higher derivative gravity in brane backgrounds.

\section{Gravity in Brane Backgrounds}
We shall begin with the following $D$-dimensional gravitational
action
\begin{eqnarray}\label{action1}
S&=&\int_{{\cal B}}
d^{D}x\,\sqrt{-g_{D}}\,\Big\{\frac{R}{\kappa_{D}}
-2\Lambda+\alpha\Big(R^2-4 R_{pq}R^{pq}
+ R_{pqrs}R^{pqrs}\Big)+{\cal L}_{m}^{bulk}\Big\}\nn\\
&{}&+\sum_k\int_{\partial{\cal
B}}^{k'th~brane}d^{D-1}\,x\sqrt{-g_{D-1}}\, \big({\cal
L}_m^{brane} -\Lambda_k\big) +\int d^d\,x
\sqrt{-g_{(d)}^{(z_1,z_2,\cdots=0)}}\,(-\T)\,,
\end{eqnarray}
where $\partial{\cal B}$ represents the $(D-1)$-dimensional
boundary. The gravitational coupling $\kappa_{D} = 16 \pi G_{D} =
M^{2-D}$, with $M$ and $G_D$ being, respectively, the $D$
dimensional mass scale and Newton constant. The indices
$(p,q,\cdots,s)=(0,1,2,3,\cdots,D-1)$, $\Lambda$ is a $D$
dimensional bulk cosmological term, $\Lambda_k\,(k=1,2,\cdots,
(D-d))$ represent the vacuum energy (brane tension) of the
$(D-2)$-branes, and $\T$ is the $d$ dimensional brane tension at a
common brane junction. For practical purposes we shall take $d=4$,
so $\alpha$ takes a mass dimension $M^{D-4}$. The second action
in~(\ref{action1}) is the effective action for $(D-2)$ branes,
while the last term introduced at the common intersection of
higher dimensional branes characterizes a four-dimensional brane
action. This term will be in effect only if $D\geq 6$, because in
$D=5$ the second action in~(\ref{action1}) has already represented
the sum of $3$-brane actions. For vacuum branes, one has ${\cal
L}_m^{brane}=0$, and also ${\cal L}_m^{bulk}=0$, since the matter
degrees are supposed to be confined on the branes.

\subsection{Choice of background and RS tunings}

Let us consider a smooth version of the multidimensional patched
$AdS$ space, introduced in Ref.~\cite{ArkaniHamed99b} to study
intersecting brane-world models, with metric \be\label{background}
ds^2=e^{-2A(z)}\left(\eta_{\mu\nu}\,dx^\mu\,dx^\nu+
\tilde{g}_{ij}\, dz^i dz^j\right)\,, \ee where
$\mu,~\nu=0,1,\cdots (d-1)$ with $N\equiv(D-d)$ being the number
of extra spatial dimensions. The Einstein field equations
($\alpha=0$) give a solution $A(z_i)=\log (\sum_{i=i}^{D-4}
|z_i|/\ell+1)$, with $\ell$ being the curvature radius of AdS
space, by satisfying the RS relations
\begin{eqnarray}\label{tuning1}
\Lambda= - \frac{(D-1)(D-2)(D-4)\, M^{D-2}}{2\ell^2}\,,\quad
\Lambda_k = \frac{4 (D-2)\,M^{D-2}}{\ell}\,.
\end{eqnarray}
One has $\T=0$ for $\alpha=0$. The four-dimensional brane tension
$\T$ at the common intersection of two $4$-branes is non-zero for
$\alpha>0$~\cite{JEKim01a,IPN01c}. The two expressions
in~(\ref{tuning1}) imply the RS fine-tuned relation \be
\label{RStuning}
\Lambda_k^2=-\frac{32\,(D-2)}{(D-1)(D-4)}\,M^{D-2}\,\Lambda\,.
\label{lambdaandT} \ee Therefore, since $\Lambda_k^2\geq 0$, the
bulk space-time is anti-de Sitter ($\Lambda<0$). For $D=5$,
$\Lambda_k$ can be replaced by $\T$, and hence $\Lambda_k$ defines
a $3$-brane tension in an $AdS_5$ space.

\subsection{Linearized Einstein gravity}

Next we consider a linearized theory. With warped space-time
metric~(\ref{background}) in the bulk, the linearized
$D$-dimensional Einstein field equations take the following form:
\begin{eqnarray}\label{linearform}
&&-\frac{1}{2}\,\partial^2 h_{pq}+ \frac{1}{2}\,\eta_{pq}\,
\partial^c\partial_c h_d^d-
\frac{D-2}{2}\,\partial^{m} A\,\Big[\partial_p h_{qm}
+\partial_q h_{p m}-\partial_{m} h_{pq}\Big] \nn \\
&&-\frac{(D-2)}{2}\Big[2\partial_{m}\partial^{m} A-
(D-3)\partial^{m}A\,\partial_{m} A
\Big]h_{pq}\nn \\
&&+\frac{(D-2)}{2}\,h_{mn}\Big[2\partial^{m}\partial^{n} A
-(D-3)\partial^{m}A\,\partial^{n}A\Big] \eta_{pq} \nn\\
&&~~=-\, \kappa_D\, \bigg[\Lambda h_{pq}\,e^{-2A(z)}+ \Lambda_k
\left(h_{\mu\nu}-\half\, \eta_{\mu\nu}\,h_m^m\right)
\delta_p^{\mu}\delta_q^{\nu}\,e^{(D-6)A(z)}\bigg]\,,
\end{eqnarray}
where we have used a harmonic gauge $\partial^q
h_{pq}=\frac{1}{2}\,h^s_s$. Since the warp factor $A(z)$ is only a
function of the extra spatial coordinates, the indices $m$ and $n$
take face values from $z_1$ to $z_{D-d}$. When $d=4$, the number
of extra dimensions $N$ is $(D-4)$. The $(\mu\nu)$ components
of~(\ref{linearform}) largely simplify in using the RS fine-tuned
relation~(\ref{tuning1}). The second term in~(\ref{linearform})
would not show up in using transverse-traceless gauge $\nabla^q
h_{pq}=0=h_p^p$, instead of the harmonic gauge.

It is more convenient to perturb the metric
background~(\ref{background}) in the following form
\be\label{background1}
ds^2=e^{-2A(z)}\left(\left(\eta_{\mu\nu}+h_{\mu\nu}\right)\,dx^\mu\,
dx^\nu+dz^i dz^i\right)\,. \ee However, in this gauge, other than
the standard graviton $h_{\mu\nu}$, there may be some additional
gravitational degrees of freedom coming from $h_{\mu m}$ and
$h_{mm}$. With more than one (conformal) extra dimensions there is
a general problem for diagonalizing the linearized fluctuations of
all metric fields, which is computationally difficult. To this
end, since a gravitational potential on the brane is mediated by
effective $4D$-gravitons, we find reasonable to analyze only the
linearized field equations for graviton $h_{\mu\nu}$ with the
gauge $h_\mu\,^\mu=0,\,\partial^\lambda h_{\lambda\mu}=0$. One may
use an approximation~\cite{ArkaniHamed99b,Csaki00a}, where the
tensor modes either decouple from the off-diagonal components of
the perturbed equations like $h_{\mu m}$ and diagonal components
$h_{mm}$, or only the nonzero components of the fluctuations are
$h_{\mu\nu}$. We comment upon the decoupling of the scalar modes
of the metric perturbations in the discussion section.

In our analysis we follow the background subtraction technique
introduced in Ref.~\cite{Csaki00a}, where one considers vacuum
branes and subtracts out the background fields from the variations
of $G_{ab}$ and $H_{ab}$ (the second order Lovelock tensor). For a
gravity action of the form~(\ref{action1}), one may define $\delta
T_{ab}=\bar{T}_{ac} h_b^c$, where $\bar{T}_{ab}
=\kappa_D\,\bar{G}_{ab}$ is taken about the background. This
approach extends in an obvious way to developing the perturbative
expansion to higher order curvature terms. An additional benefit
of this approach is that the RS fine tunings of the previous
subsection will be only implicit. Furthermore, it is reasonable to
impose the gauge $\partial_\mu h^{\mu\nu}=0$, $h_\mu^\mu=0$. With
these approximations, the Einstein equations $\delta
G_{ab}=\kappa_{D}\delta T_{ab}$ linear in $h_{\mu\nu}$ take a
remarkably simple form~\cite{Csaki00a} \be -\,\partial_a\partial^a
h_{\mu\nu}+(D-2)\,\partial^m A\,\partial_m h_{\mu\nu}=0\,. \ee We
define $h_{\mu\nu}=e^{(D-2)A(z)/2}\,\tilde{h}_{\mu\nu}$ and
$\Box_4\,\tilde{h}_{\mu\nu}=m^2 \tilde{h}_{\mu\nu}$, where
$\tilde{h}_{\mu\nu}=\epsilon_{\mu\nu} e^{ip.x}\psi(z)$,
$\epsilon_{\mu\nu}$ is the polarization tensor, and arrive at the
following analog non-relativistic Schrodinger equation
\begin{equation}
\left(-\partial_{z_i}^2+\frac{D(D-2)}{4\big(|z_i|+\ell\big)^2}
\,\sum_{i=1}^{D-4}sgn(z_i)^2-\frac{(D-2)}{\left(|z_i+\ell|\right)}\,
\delta(z_i)\right)\psi(z_i)=
m^2\psi(z_i)
\label{rsequation}\,.
\end{equation}
We are considering here the case where warp factor $e^{-A(z)}$ is
a function of all transverse coordinates $z_i$, where
$i=1,2,\cdots,(D-4)$, so that $\sum_i^{D-4}sgn(z_i)^2=(D-4)$ other
than at the origin (brane-junction) in the transverse space. In
order to normalize four-dimensional metric at the origin, one has
to assume that $A(0)=0$. For $D=5$, Eq.~(\ref{rsequation}) gives
the RS one-dimensional Schr\"odinger equation, and this was
extensively studied, for example, in
Refs.~\cite{Garriga99a,Giddings00a,Csaki00a}. So we shall be
interested only in the $D=6$ case.

Define a set of new coordinates $x_{\pm}\equiv |z_1|\pm |z_2|$, so
that the bulk part of~(\ref{rsequation}) takes the form \be
\bigg[-\partial_{x_-}^2-\partial_{x_+}^2
+\frac{6}{(|x_+|+|x_-|+\ell)^2}\bigg]\hat{\psi}(x_-,x_+)= \half
\,m^2\, \hat{\psi}(x_-,x_+)\,,\label{uveqn} \ee where
$\hat{\psi}(x_-,x_+)\equiv \psi(z_1,z_2)$. We can separate this
equation into \be
-\partial_{x_-}^2\varphi(x_-)=m_-^2\,\varphi(x_-)\,,\quad {\mbox
and} \quad
\left(-\partial_{x_+}^2+\frac{6}{(|x_+|+\ell)^2}\right)\varphi(x_+)
=m_+^2\,\varphi(x_+)\,, \ee where we have defined
$m^2=2\big(m_-^2+m_+^2\big)$, and
$\hat\psi_m=\varphi_{m_-}(x_-)\times \varphi_{m_+}(x_+)$. Since
$m^2>0$, $m_-^2$ and $m_+^2$ each can have only positive eigen
energy. The solution for each continuum wavefunction is a linear
combination of Bessel functions \be \varphi_{m_-}(x_-)=
N_1(m_-)\sqrt{|x_-|}\, \left(\sin(m_- x_-)+A_m\cos(m_-
x_-)\right)\,. \ee \be \varphi_{m_+}(x_+) =
N_2(m_+)\,\sqrt{(|x_+|+\ell)}\, \left[J_{5/2}\left(m_+
(|x_+|+\ell)\right) +B_m\,Y_{5/2}\left(m_+
(|x_+|+\ell)\right)\right]\,. \ee These solutions must satisfy the
Neumann type boundary conditions at the brane junction
$x_{\pm}=0$: \be \left(x_{\pm}\,\frac{\partial_{x_+}\pm
\partial_{x_-}}{2}+\frac{5}{2} \right) \psi(x_\pm)=0\,. \ee The
coefficients $A_m$ and $B_m$ can easily be read off \be A_m=\cot
(m_- x_-)\,, \quad B_m=-\frac{Y_{3/2}(m_+ \ell)}{J_{3/2}\big(m_+
\ell)}\,. \ee Thus $\psi_{m_+}(0)\sim (m_+\ell)^{\alpha-1}$, where
$\alpha+1/2=5/2$. There are $4$ $(=2^{D-4})$ sectors in the mass
spectrum spanned by (i) $\varphi_0(x_-)\,\varphi_0(x_+)$, (ii)
$\varphi_{m_-}(x_-)\,\varphi_0(x_+)$, (iii) $\varphi_0(x_-)\,
\varphi_{m_+}(x_+)$, and (iv) $\varphi_{m_-}(x_-)\,
\varphi_{m_+}(x_+)$. The state $(i)$ gives the four-dimensional
graviton, and the set of continuum states $(ii)$ or $(iii)$, which
is localized in $x_-$ or $x_+$ direction, contributes as an
integral over the single eigenvalue $m_-$ or
$m_+$~\cite{Csaki00a}. The set of continuum states $(iv)$
contributes to Newton's law. For two point masses $m_1$ and $m_2$
placed at a distance $|x-x'|=r$ on the four-dimensional brane
intersection, the Newtonian potential is \be -\Delta_4 V(r)\simeq
\frac{m_1\,m_2}{M_{(6)}^{4}}\,\frac{1}{\ell}\, \int_{m_0}^{\infty}
dm_+\,\left[\int_{m_0}^{\infty} dm_-\,
\frac{e^{-\widehat{m}\,r}}{r}\,
 {|\,\varphi_{m_-}(0)|}^2 \right]\,{|\,\varphi_{m_+}(0)|}^2\,,
\ee where $\widehat{m}=\sqrt{m_-^2+m_+^2}$. A qualitative feature
of this potential can be known by specializing to the case where
$m_-$ would extend down to $m_0=0$, and $\psi_{m_-}(0)=1$. Hence
\be - \Delta_4 V(r)\simeq G_N^{(4)}\,\frac{m_1\,m_2}{r}\,c_1
\left(\frac{\ell}{r}\right)^3 \,, \ee where $1/G_N^{(4)}\sim
\ell^2 M_{(6)}^4$, $c_1$ is a number of order one. For large $D$
and $\ell<r$, the Kaluza-Klein mode corrections
$\left(\ell/r\right)^{D-3}$ are more suppressed, so gravity
becomes weaker as the number of transverse directions increases.
Similar results are known for gravity localized on a
four-dimensional string-like defect in $D=6$~\cite{Gherghetta00a},
a difference here is that now we have two (non-compact) transverse
directions, and this is itself more interesting in the scenario of
Refs. ~\cite{Dvali98a} and \cite{ArkaniHamed99b}.

\section{Corrections to Einstein Gravity}

Next we investigate the brane-world solutions which occur due to a
non-trivial Gauss-Bonnet coupling $\alpha$. The exact metric
solution for the modified Einstein equations is given by
$A(z)=\log (\sum_k|z_k|/L+1)$, where $L$ is the AdS curvature
radius which has a contribution from the GB coupling. We should
note that $L^2$ is defined by the bulk solution~\cite{IPN01c} \be
\label{Lsquare}
\frac{1}{L^2}=\frac{1}{2(D-4)^2(D-3)\,\alpha\,\kappa_{D}}
\left[1\pm\sqrt{1+\frac{8(D-3)(D-4)\,
\alpha\Lambda\,\kappa_{D}^2}{(D-1)(D-2)}}\right]\,. \ee The
space-time metric is therefore
\begin{equation} \label{finalbg}
ds_{D}^2= \frac{1}{(\sum_{i=1}^n\,|z_i|/L+1)^2}\,
\Big(\eta_{\mu\nu}dx^{\mu}dx^{\nu} +\sum_{j=1}^n\,(d z_j)^2\Big)\,.
\label{conmetric2}
\end{equation}
One may view the above metric background as a four-dimensional
intersection of higher dimensional branes or as a $3$-brane
embedded in space of dimensions $D\geq 5$. One has
$L^2=2(D-4)^2(D-3)\,\alpha\,\kappa_{D}$ when the two branches of
the bulk solution~(\ref{Lsquare}) coincide. The RS type background
relations
\begin{eqnarray}\label{tuning2}
\Lambda= - \frac{(D-1)(D-2)(D-4)\,(1-\gamma/2)}{2L^2\,\kappa_{D}}\,,\quad
\Lambda_k = \frac{4(D-2)\,(1-\gamma/3)}{L\,\kappa_{D}}\,,
\end{eqnarray}
may be assumed implicitly for $\gamma<1$. Here
$\gamma=2(D-3)(D-4)^2\,\alpha\,\kappa_{D}/L^{2}$ and $\alpha$ is
the usual Gauss-Bonnet coupling, so $\gamma=0$ when $\alpha=0$. In
this limit, one finds from~(\ref{tuning2}) the RS fine-tuned
relations. We may define $\alpha=M^{D-5}\,\alpha_1$, with
$\alpha_1$ being a dimensionless GB coupling. If we define
$l=L/\sqrt{(D-4)(1-\gamma/2)}$, then $\Lambda$ takes the usual
form $\Lambda=-\,(D-1)(D-2)/(2\,\kappa_D\,l^2)$.

Let us include a matter source $T_{\mu\nu}^{(m)}$ on the brane,
and study linearized equations for the effective four-dimensional
gravitational fluctuations. We consider only the scalar wave
equation for each of the components $h_{\mu\nu}$ in the
background~(\ref{finalbg}). The linearized field equations for
$h_{\mu\nu}$ read
\begin{eqnarray}\label{tt-part}
&{}&\bigg[1-\frac{2(D-3)(D-4)^2\,\alpha\kappa_D}{L^2}\bigg]\,
\Big(-\partial_\lambda^2-\partial_z^2+\frac{(D-2)}{L}\,sgn(z_i)
\partial_{z_i}\Big)h_{\mu\nu}(x,z)\nn \\
&{}& +\,\frac{2(D-4)\,\alpha\kappa_D}{L} \sum_{i\neq j}
\delta(z_i)\Big[-\partial_\lambda^2-\partial_{z_j}^2+\frac{(D-3)}{L}\,
\sum_k sgn(z_k)\partial_{z_k}\Big]\, h_{\mu\nu}(x,z)=- \kappa_D\,
T^{(m)}_{\mu\nu}\,.\label{4dttgraviton}
\end{eqnarray}
One of the important features of the modified solutions due a
non-trivial $\alpha$ is that a Gauss-Bonnet term contributes with
a delta-function term to the linearized equations, which implies a
non-trivial topology at the brane junction, like that a non-zero
four-dimensional brane tension at the common brane-intersection
with several co-dimension one branes. In Eq.~(\ref{tt-part}),
terms involving Dirac delta functions vanish for $|z|>0$.
Therefore, for $z\neq 0$, one has a RS type bulk equation but
multiplied by the factor
$(1-2(D-3)(D-4)^2\,\alpha\,\kappa_{D}/L^{2})=1-\gamma$. This
apparently implies that the effect of a GB term may be removed by
redefining the Einstein constant. But this is not the case, rather
a non-trivial GB coupling $\alpha$ alters the physics of the
brane-world by modifying the Neumann propagator in the bulk. If
one allows the parameters to take values such that the first
square bracket term in~(\ref{tt-part}) vanishes, the solutions
become solitonic. This is because what survives after this setting
precisely gives a RS equation multiplying with an additional delta
function term, but in one co-dimension lower. As in the RS
solution, which is given by $\gamma=0$, there always exist
brane-world solutions for $0<\gamma \leq 1$. So the physical
relevance of the GB coupling $\alpha$ is two fold. If one has
solutions satisfying $0<\gamma<1$, such solutions will be fairly
similar to the RS solutions except with some corrections, like in
the Newtonian potential. But if one is allowed to take $\gamma\sim
1$, one finds a new branch of solution with finite effective
gravitational constant without finite distance bulk singularity.
In the next section, we shall be interested in the $\gamma<1$
solution.

\section{Green Function in $D$ Dimensions}
In this section we analyze the Green functions by expressing a
$D$-dimensional propagator $\G_{D}(x,z;x^\p, z^\p)$ in terms of
the Fourier modes such that
\begin{eqnarray}
\G_{D}(x,z;x^\p, z^\p)=\int\frac{d^{D-1} p}{(2\pi)^{D-1}}\,
e^{ip(x-x^\prime)}\G_p(z,z^\p),\label{fourier1}
\end{eqnarray}
where the Fourier components $\G_p(z,z')~(\equiv \tilde{h}(z,z')
\sim e^{ip.x}\psi(z,z'))$ satisfy
\begin{equation}
\left(1-\gamma\right)
\Big(\partial_z^2 - p^2-\frac{D-2}{z}\,\partial_z\Big)\G_p(z,z^\prime)
=e^{(D-2)A(z)}\, \delta(z-z^\prime)
\label{propa1}\,.
\end{equation}
When one of the arguments is at $z'=L$, the Neumann propagator is
calculated in Appendix D to be
\begin{equation}
\G_{D}(x,z;x',L)=(1-\gamma)^{-1}\left(\frac{z}{L}\right)^{\nu}
\int\frac{d^{2\nu} p}{(2\pi)^{2\nu}}\, e^{ip(x-x')}\,\frac{1}{q}
\bigg[\frac{H^{(1)}_{\nu}(qz)}{H^{(1)}_{\nu-1}(qL)
+\chi\, (q L)\, H^{(1)}_{\nu}(qL)}\bigg]
\label{greengen}\,,
\end{equation}
where $\nu=(D-1)/2$, $\chi=2\gamma/((D-3)(1-\gamma))$, and
$H^{(1)}=J+iY$ is the first Hankel function and $q^2=-p^\mu
p_\mu=m^2$. For both the arguments of the propagator at $z, z'= L$
(on the brane), using the Bessel recursion relations, one can find
a two point correlator of the effective theory. For example, when
$D=5$, a two-point correlator is \be \langle
\phi(\vec{p})\phi(-\vec{p})\rangle =
\frac{1}{(1+\gamma)}\left[\frac{2}{q^2 L}-\frac{1}{q}\,
\frac{(1-\gamma)\,H_0^{(1)}(qL)}
{(1+\gamma)\,H_{1}^{(1)}(qL)-\gamma\,H_0^{(1)}(qL)}\right]\,. \ee
For arbitrary $D$, by using Bessel expansions in
Eq.~(\ref{greengen}), we obtain the scalar Neumann propagator
\begin{equation}\label{greensum}
\G_{D}(x,L;x', L)=(1+\gamma)^{-1}
\bigg[\frac{(D-3)}{L}\, \G_{D-1}(x,x^\prime)
+\G_{KK}(x,x')\bigg]\,,
\end{equation}
where
\be
\G_{D-1}(x,x^\prime)=\int\frac{d^{D-1} p}{(2\pi)^{D-1}}
\frac{e^{ip(x-x^\prime)}}{q^2}\,,
\label{zeromode}
\ee
\bea
\G_{KK}(x,x^\prime) &\simeq& -\int\frac{d^{D-1} p}{(2\pi)^{D-1}}\,
e^{ip(x-x^\prime)}\,
\bigg[\frac{\left(D-4+(D-6)\gamma\right)}
{2(D-5)}\frac{1}{(1+\gamma)}\,qL \nn \\
&{}& + \frac{1-\gamma}{1+\gamma}\,\frac{(qL)^{D-4}}{q\,C_1}\,
\ln\left(\frac{qL}{2}\right)\bigg]\,,
\label{kkmode}
\end{eqnarray}
where $C_1$ is a dimension dependent number (see App. D). For
$\gamma=0$, the above results reproduce the scalar Neumann
propagator found in Ref.~\cite{Giddings00a}, after a substitution
$d=D-1$, and the results in Ref.~\cite{IPN01b} with $D=5$. The
first term in~(\ref{greensum}) is the standard propagator of a
massless scalar field.

The long distance behavior, $r>>L$, of the propagator is governed
by a small $q$ behavior of the Fourier modes. Thus, for $qL<<1$, a
leading order contribution to the propagator comes from the
logarithm. For $D=5$, and $|x-x'|>>L$, $qL<<1$, we find, to the
leading order in $q$,
\begin{equation}
\G_4(x,x')= \int\frac{d^4p}{(2\pi)^4}\,\frac{e^{ip(x-x')}}{q^2}
\propto \frac{1}{|x-x'|^{2}}\label{g4delta4}\,.
\end{equation}
Thus $\G_4(x,x')$ is just the ordinary massless scalar propagator
in four-dimensions. Eq.~(\ref{greensum}) implies that a
four-dimensional massless graviton propagator is contained in the
full five-dimensional propagator. This is one of the plausible
results of the RS2 model supplemented by a GB term. By the same
token, for the KK modes, one has \bea
\G_{KK}(x,x')&\simeq&\frac{1-\gamma}{1+\gamma}\,L^{D-4}\,
\int\frac{d^{D-1} p}{(2\pi)^{D-1}}\,
e^{ip(x-x^\p)}\,q^{(D-5)}\,\ln( qL/2)\nn\\
&\propto& \frac{1-\gamma}{1+\gamma}\,
\frac{L^{D-4}}{|x-x'|^{2D-6}}\,.\label{kkmode1} \eea For
$1>\gamma>0$, we expect some correction in Newton's law. At large
distance scales on the brane, $|x-x'|=r>>L$, the contribution of
the KK modes~(\ref{kkmode1}) is very small compared to the zero
mode contribution~(\ref{greensum}). Like the $\gamma=0$ solution
~\cite{Giddings00a}, for $D\geq 5$ and odd $D$, a leading behavior
of $\G_{KK}$ goes like $\sim r^{6-2D}$. When $D$ is even, there
are no logarithmic terms and a leading behavior of $\G_{KK}(x,x')$
goes like $\sim L\,r^{1-D}$, which is further more suppressed
compared to the zero-mode contribution. Therefore, the effect of
gravity becomes weaker when the number of extra (transverse)
dimensions grows. One also notes that the Newtonian potential for
a point source of mass $m$, and $D=5$, is \be V(r)\simeq
-\,\frac{G_4
m}{r}\left[1+\left(1-\frac{2\gamma}{1+\gamma}\right)\frac{1}{2}\,
\frac{L^2}{r^2}\right] \,.\ee In this formula, the contribution of
the brane-bending mode is not included, which might change the
factor $1/2$ into $2/3$~\cite{Garriga99a,IPN01b}. It is seen that
the GB term contributes to Newtonian potential with the opposite
sign to that of the Einstein term or scalar curvature. One can
expect a small correction to Newton' law with $\gamma<1$ and
$r>L$, but such a correction is almost trivial for $r>>L$ or/and
$\gamma\sim 1$.

\section{Nonperturbative Solutions with a GB Term}

A non-singular Minkowski or de Sitter brane-world compactification
could be difficult for Einstein's
theory~\cite{Freedman99a,Malda00a}. So one is lead naturally to
hope that adding higher derivative corrections to the brane-world
action might improve this situation. As a convenient choice, we
shall begin with the Lagrangian of gravity, including a GB term
and a bulk scalar field, in the form \be \label{thick1} S=\int
d^{D}x\,\sqrt{-g_{D}}\,\left(\frac{R}{\kappa_{D}} +
\alpha\left(R^2-4 R_{pq}R^{pq} +
R_{pqrs}R^{pqrs}\right)-\frac{1}{2}\,(\partial \varphi)^2 -2
V(\varphi) +\cdots\right)\,. \ee The dots in~(\ref{thick1})
represent higher order terms in the scalar field $(\varphi)$ and
some other fields which are not turned on. We assume that
$V(\varphi)$ is non-positive (bulk) cosmological potential, and
$\varphi$ can only depend on extra coordinates,
$\varphi=\varphi(z)$, as dictated by the Poincar\'e invariance on
a brane. Though we will not restrict the action~(\ref{thick1}) to
a particular string theory background, one may think about it like
including $\alpha^\p$ corrections in type I string theories. In
propagator correction-free Gauss-Bonnet scheme~\cite{Metsaev87a},
the $R^2$-corrections can have a dependence on a scalar field,
$\alpha\to \frac{\alpha^\p}{2}\, \lambda_0 e^{-\,m\varphi}$, with
$m=1/\sqrt{D-2}$, $\lambda_0=1/4~(1/8)$ for bosonic (heterotic)
string. For a constant bulk scalar, since $V(\varphi)$ takes a
bare value $\Lambda_0$, this parameterization is not so important
for our analysis.

We can write the metric background in the form \be
\label{cmetric1} ds_D^2=\Omega^2(z)\left(dx_d^2+\widehat{g}_{mn}
dz^m dz^n\right)\,. \ee Here $\Omega(z)=e^{-A(z)}$,
$dx_d^2=\tilde{g}_{\mu\nu}\,dx^\mu dx^\nu$ with $\tilde{g}$ being
the $d$-dimensional metric, which can be Minkowski, de-Sitter or
anti-de Sitter space. We study only the case of $D$-dimensional
warped geometry compactified to $d$-dimensional RS type space-time
($\tilde{g}_{\mu\nu}=\eta_{\mu\nu}$).

The equations of motion in $D$-dimensions take the form \be
\label{nogo1} R_{ab}(g)
=\kappa_D\left(\tau_{ab}-\frac{1}{D-2}\,g_{ab} \tau_c^c\right)
+\kappa_D\,{\cal T}_{ab}\equiv \kappa_D\,T_{ab}\,, \ee where the
contribution to the stress energy of a massless scalar field and a
bulk cosmological potential reads \be \tau_{ab}=\partial_a
\varphi\,\partial_b \varphi -\frac{1}{2}\,g_{ab}
(\partial\varphi)^2-V(\varphi) g_{ab}\,, \ee and, the contribution
of the GB interaction term reads \be {\cal
T}_{ab}=\frac{\alpha}{D-2}\,g_{ab} {\cal R}_{GB}^2-2\alpha\left(R
R_{ab}-2 R_{acbd}R^{cd}+R_{acde}R_b\,^{cde} -2R_a\,^c
R_{bc}\right)\,. \ee

\subsection{ Energy conditions with a Gauss-Bonnet term}

Let us take the $(\mu\nu)$ components of the $D$ dimensional Ricci
tensor \be R_{\mu\nu}=R_{\mu\nu}(\tilde{g})-\tilde{g}_{\mu\nu}
\left(\widehat{\nabla}^2\log\Omega+(D-2)(\widehat{\nabla}\log\Omega)^2
\right)\,. \ee In the RS compactification, $d$-dimensional space
is a Minkowski space, so $R(\tilde{g})=0$. In our metric
convention (mostly positive), for $\alpha=0$, the "c-theorem" of
Ref.~\cite{Freedman99a} reads
\begin{eqnarray}
-(d-1)\,e^{2A(z)}
\left(A^{\prime\prime}+{A^\prime}^2\right)&=&{R}_t^t-{R}_z^z
=\kappa_{d+1} \left({\tau}_t^t-{\tau}_z^z\right)\nn\\
&=&
-\kappa_{d+1}\, e^{2A(z)}\left(\partial\varphi\right)^2\leq 0 \,,
\end{eqnarray}
where $V(\varphi)$'s contribution to $\tau_t^t$ and $\tau_z^z$
cancels out. This is the obvious positive energy condition
$\tau_{ab}\,\xi^a\xi^b\geq 0$, where $\xi^a$ is any arbitrary
future-directed timelike or null vector. Since \bea
\label{munuandzz} {\cal T}_{\mu\nu}&=&(d-2)(d-3)\alpha
\left(2A^{\p\p}-(d-2){A^\p}^2\right)
{A^\p}^2 e^{2A(z)}\,\eta_{\mu\nu}\,,\nn \\
{\cal T}_{zz}&=&d(d-2)(d-3)\alpha \left(2A^{\p\p}+{A^\p}^2\right)
{A^\p}^2 e^{2A(z)}\,, \eea there arises an identical condition
from \be \kappa_{d+1}\,\left({\cal T}_z^z-{\cal
T}_t^t\right)=2\varepsilon (d-1) e^{2A(z)}{A^\p}^2
\left(A''+{A'}^2\right)e^{2A(z)}\geq 0 \,, \ee where
$\varepsilon=(d-2)(d-3)\alpha\kappa_{d+1}$. The condition
$A^{\p\p}+{A^\p}^2\geq  0$ may be equivalent to the weak energy
condition (WEC) $-\rho +|p_i|\geq 0$, with $\rho$ and $p$ being
the energy density and pressure. For $\alpha>0$, the strong energy
condition $\left(\tau_{ab}-\frac{1}{D-2}\,g_{ab} \tau\right)\xi^a
\xi^b\geq 0$ is promoted to ${T}_{ab}\,\xi^a \xi^b\geq 0$. First
consider \bea \label{strong1}
\left(\tau_{ab}-\frac{1}{D-2}\,g_{ab}\,\tau\right)\,\xi^a\xi^b
&=&\left(\partial_a\varphi\,\partial_b\varphi-\frac{2}{D-2}\,
V(\varphi)\,g_{ab}\right)\xi^a\xi^b \nn \\
&=&\left(\partial_z\varphi\right)^2\,\xi^z\xi_z- \frac{2}{D-2}\,
V(\varphi)\left(\xi^\mu\xi_\mu+\xi^z\xi_z\right)\geq 0 \,. \eea So
the strong energy condition is not violated if $\xi^a$ is a null
vector and $V(\varphi)<0$. For a future-directed null vector,
$\xi^a\xi_a=e^{2A(z)}\left(\xi^\mu\xi_\mu+\xi^z\xi_z\right)=0$,
Eq.~(\ref{strong1}) implies that
$-\left(\partial_z\varphi\right)^2\xi^\mu\xi_\mu\geq 0$, so
$\xi^\mu$ can be timelike ($\xi^\mu\xi_\mu=-1$) or null-like
$(\xi^\mu\xi_\mu= 0)$. This is consistent with the condition
$R_{ab}\xi^a\xi^b\geq 0$: \be \label{Rabxiab} -
(d-1)\,e^{2A(z)}\left({A^\p}^2+A^{\p\p}\right)\xi^\mu\xi_\mu\geq 0
\,. \ee We would like to see whether the second piece
in~(\ref{nogo1}), i. e., the contribution to stress energy from
the GB term, violates the strong energy condition. For a null
vector $\xi^a$, the condition ${\cal T}_{ab}\,\xi^a\xi^b\geq 0$,
using~(\ref{munuandzz}), yields \be
\left(-2\alpha\,(d-1)(d-2)(d-3)
\left({A^\p}^2+A^{\p\p}\right)e^{2A(z)}{A^\p}^2\right)
e^{2A(z)}\,\xi^\mu\xi_\mu\geq 0 \,. \ee Thus for all non-spacelike
$\xi^\mu$, i. e., $\xi^\mu\xi_\mu\leq 0$, the strong energy
condition (SEC) is intact provided that the WEC
$A^{\p\p}+{A^\p}^2\geq 0$ holds. This may not be the case for any
other combination of Riemann tensors or $R^2$ terms~\footnote{We
acknowledge fruitful correspondences with C. Nu\~nez about the
no--go theorem and energy conditions with a GB term that prompted
us to add the above explanation.}. Here, validity of the SEC
basically says that brane gravity is attractive.

\subsection{Randall-Sundrum limit}

Let us introduce a brane action for the RS singular $3$-brane with
a positive brane tension $\sigma>0$: \be \label{braneaction}
S_{brane}=2\int d^4x\,\sqrt{|g_4|}\, (-\sigma)\,. \ee For a
constant scalar, the field equations following from~(\ref{nogo1})
in $D=5$, including a contribution from ~(\ref{braneaction}),
simplify to
\begin{eqnarray}
3\left({A^\p}^2+A^{\p\p}\right)&=&
3\left({A^\p}^2+A^{\p\p}\right) \left(2\varepsilon\,
e^{2A(z)}{A^\p}^2\right)+\sigma\,
\kappa_5\, \delta(z)\label{constphi1}\\
12{A^\p}^2&=&-\,\kappa_5\,\Lambda_0\,e^{-2A(z)}+6\left(2\varepsilon\,
e^{2A(z)}{A^\p}^2\right) {A^\p}^2 \,, \label{constphi2}
\end{eqnarray}
where $\varepsilon\equiv 2\alpha\kappa_5$. For
$\varepsilon=0$, the bulk equation $A^{\p\p}+{A^\p}^2=0$ may imply
that $e^{2A(z)}{A^\p}^2~(\equiv C)$ is a constant. However, $C$ is
already fixed by Eq.~(\ref{constphi2}) such that
$C=-\,\frac{\kappa_5\Lambda_0}{12}\equiv \frac{1}{\ell^2}$, where
$\ell$ is a constant with dimension of length. Solving the
equation $A^{\p\p}+{A^\p}^2=0$ amounts to selecting a solution
$A^\p(z)=\left(|z|+z_*\right)^{-1}$ of the unperturbed Einstein
theory, but $z_*$ is undetermined by the bulk equations. One may
fix $z_*$ using the continuity condition and normalization
condition, such that $A^\p(0)=1/\ell$, and arrive
at~\cite{Randall2} \be \label{RStunings}
\sigma=\frac{6}{\kappa_5\,\ell}\,,\quad
\Lambda_0=-\,\frac{12}{\kappa_5\,\ell^2}\,. \ee As is well known
one has some undesired features with $A^{\p\p}+{A^\p}^2=0$ as an
initial condition. For example, there is a finite-distance bulk
singularity for $z_*<0$ at $z=z_*$. So we will now investigate the
case of interest, $\varepsilon>0$. In this case, the bulk equation
can be satisfied even by selecting \be 2\varepsilon
e^{2A(z)}{A^\p}^2= 1\,.\label{secondsol} \ee It is obvious that
this freedom is not there with $\varepsilon=0$. Since
$\varepsilon>0$, the metric solution is \be \label{solution}
 e^{A(z)}\,=\, \int_0^{z}\frac{1}{\sqrt{2\varepsilon}}\,dz \quad
\Rightarrow  e^{-A(z)}\,=\,
\frac{\sqrt{2\varepsilon}}{|z|+\sqrt{2\varepsilon}}\,, \ee where
we have normalized the solution $A(0)=0$, such that $e^{-\,A(z)}$
takes a value $1$ at $z=0$. We find it convenient to define the
length scale $\sqrt{2\varepsilon}=l$, so
$A^{\p\p}+{A^\p}^2=\frac{2\delta(z)}{(|z|+l)}$. The
condition~(\ref{secondsol}) therefore, unsurprisingly, also the
solves the bulk equation $A^{\p\p}+{A^\p}^2=0$. It is obvious that
$A^{\p\p}+{A^\p}^2=\frac{2}{l}$ on the brane ($z=0$), as it should
be in order to keep the WEC $A^{\p\p}+{A^\p}^2\geq 0$ intact. The
solutions with $\varepsilon>0$ have all essential properties of
the RS solutions, such as $e^{-A(z)}$ converges as $z\to
\pm\,\infty$, and there is no any finite distance bulk
singularity. However, there is now a new length scale in the
problem, $l$, and a common feature of these new solutions is that
they are not analytic in the coefficient $\alpha$ of the
Gauss-Bonnet interaction, so $\varepsilon>0$ is a physical
requirement. Remarkably, the RS solution with $\varepsilon=0$
corresponds to a limit of these solutions where the singularities
are pushed to infinity, $z\to \pm \infty$, so are harmless as they
might have interpretations in field theory as (ultraviolet) energy
cut-off scale.

The boundary condition relates $l$ to the brane tension $\sigma$.
For a brane at $z=0$ with positive tension, $\sigma>0$,
Eq.~(\ref{constphi1}) implies that \be
\frac{6\delta(z)}{l+|z|}\,\left(1-\frac{2\varepsilon}{l^2}\,
sgn(z)^2\right)=\sigma\,\kappa_5\,\delta(z)\,. \ee This after
regularizing the $\delta$-function,
$\delta(z)\,sgn(z)^2=\delta(z)/3$, determines the brane tension
\be \label{tension1} \sigma=\frac{1}{\kappa_5}\,\frac{4}{l}\,. \ee
The tension $\sigma$ is generally not fine-tuned because $l$ is
arbitrary. However, the scale $l$ is used also to fix the bulk
cosmological constant, via Eq.~(\ref{constphi2}), \be
\Lambda_0=-\,\frac{1}{\kappa_5}\,\frac{6}{l^2}\,. \ee There is a
fine-tuning between $\sigma$ and $\Lambda_0$, which is required to
maintain flatness of the $3$-brane. This may be relaxed for more
general solutions, like (anti-) de Sitter branes, with
$\varepsilon>0$.

A remark is in order. The bulk scale $l$ is somehow fixed by the
coefficient $\varepsilon$, but this is not an unnatural choice
from the view point of low energy effective string action, rather
a common result in higher-curvature stringy gravity, see, for
example, Refs.~\cite{Zee,Mavro,IPN00}. In the Einstein frame, the
tree-level (bosonic) string action, with appropriate conformal
weights for a dilaton $\phi$, reads~\cite{Metsaev87a,Rizos02a}
\bea \label{thick2} S_{bulk}&=&\frac{1}{\kappa_D}\int
d^{D}x\,\sqrt{-g_{D}}\,\Bigg(R +
\lambda_0\alpha^\p\,e^{-\,m\phi}\left({\cal
R}_{GB}^2+m^2\,\frac{D-4}{D-2}\,
\left(\partial\phi\right)^4\right)\nn \\
&{}& -\frac{2(D-10)}{3\alpha^\p}\,e^{m\phi} -\,m\,(\partial
\phi)^2+{\cal O}({\alpha^\p}^2)\Bigg)\,,\eea where $m=4/(D-2)$.
Therefore, with $\phi=\phi_0=const$ and $D<10$, the bulk
cosmological term $\Lambda_0\propto -\,1/\alpha^\p\sim
-\,1/\varepsilon$. The above action diverges at $\alpha^\p=0$, so
one might require $\alpha^\p>0$, so $\varepsilon>0$, in order to
get solutions which are free of singularities. In the context of
lowest-order brane-world Einstein gravity, so $\alpha^\prime=0$,
there arise naked bulk singularities due to the fact that both
dilaton and graviton field exhibit logarithmic
singularities~\cite{Sundrum00a}. This problem can be resolved by
considering the leading order $\alpha^\prime$ corrections as
recently shown in Ref.~\cite{Rizos02a}.

\section{Brane-World No--Go Theorem}

Basically, with warped space-time metrics in the bulk, there are
three different arguments given for the brane-world
``no--go theorem'' in the Einstein theory. They are \\
$\bullet$ No singularity free solution with a finite Planck mass
is possible without a fine tuning~\cite{CEGH}. In other words,
singularities in the self-tuned solutions are generic if gravity
is to be
localized. \\
$\bullet$ It is impossible to have
$e^{A(z)}A^\p(z)$ approach a positive constant as $z\to \infty$
and a negative constant as $z\to -\,\infty$. This monotonicity of
$e^{A(z)}A^\p(z)$ is often called the brane-world
$c$--\mbox{theorem}~\cite{Freedman99a} ~\footnote{In terms of the
$y$ coordinate such that $dy=e^{-\,A(z)} dz$, one has
$A^\p(y)=e^{A(z)}\,A^\p(z)$, $A^{\p\p}(y)=e^{2A(z)}
\left(A^{\p\p}(z)+{A^\p}^2(z)\right)$.}.\\
$\bullet$ There are no non-singular Randall-Sundrum or de-Sitter
compactifications where the only possible singularities occur when
the warp factor $e^{-\,A(z)}$ goes to zero at the
singularity~\cite{Malda00a}.

The first argument above is mainly related to the fine tuning of
the cosmological constant, which may complement the
Hawking-Penrose singularity theorem~\cite{Hawking70a}, rather than
being intrinsic to the Randall-Sundrum models~\footnote{In the RS
brane-world context, this implies that the generic initial
conditions, such as $A^{\p\p}(z)+{A^\p(z)}^2=0$ for $z> 0$, lead
to singular solutions of Einstein equations.}. It is interesting
to know whether any of these arguments can be avoided. It should
be possible to circumvent some of these arguments by considering
higher derivative corrections to the gravity
equations~\cite{Malda00a}.

\subsection{No--go theorem and possible avoidance}

Let us take the $(\mu\nu)$ components of ~(\ref{nogo1}) and
contract the $\mu\nu$-indices to arrive at \begin{eqnarray}
\label{fullRmumu}
d\left[(d-1){A'}^2-A''\right]&=&\kappa_{d+1}e^{-2A(z)}\,\tilde{T}\nn
\\
&{}& -\,d(d-2)\alpha\,\kappa_{d+1}
\left[2A^{\p\p}-(d-2){A^\p}^2\right] e^{2A(z)}{A^\p}^2\,,
\end{eqnarray}
where \be \label{energycond1} \tilde{T}\equiv
-\,\tau_\mu^\mu+\frac{d}{d-1}\,\tau_c^c =-\,\frac{2d}{d-1}\,
V(\varphi)\geq 0\,. \ee Here we have used the fact that
$\varphi=\varphi(z)$, and assumed $V(\varphi)<0$.

First we briefly review the no-go theorem advocated in
Ref.~\cite{Malda00a}. With $\alpha=0$ and $d=4$, one has  \be
\label{r-mumu}
\Omega^{3}\hat{\nabla}^2\Omega^{3}=3\,e^{-\,6A(z)}\,
\left[3{A'}^2-A''\right]\geq 0\,. \ee Integrating~(\ref{r-mumu})
over the compact internal space by parts one finds $\int dz
\sqrt{\hat{g}}\, (\hat{\nabla}e^{-3A(z)})^2\leq 0$. This may be
satisfied only if one allows the equality sign in~(\ref{r-mumu})
and $e^{-A(z)}$ is a constant. The condition $R_{00}=\tau_{00}-
\frac{1}{D-2}\,g_{00}\,\tau_c^c=3{A^\prime}^2-A^{\p\p}=0$, or
$R_{00}=\tilde{T}=0$, may not generate gravitational fields in the
bulk space-time~\cite{Gibbons}. Moreover, in the RS
compactification $e^{-A(z)}$ should not be a constant, rather this
factor is essential to explain the warped nature of a bulk
geometry and hence a RS compactification.

We will now investigate the case of interest, $\alpha>0$, so
$\varepsilon>0$. For $d=4$, the four-dimensional parts of the
curvatures read \begin{eqnarray}
R(x)&=&
R^{(4)}(x)+4\,e^{2A(z)}\left(2A^{\p\p}-(N+2){A^\p}^2\right)\nn
\\
{\cal R}_{GB}^2(x)&=&{{\cal R}^2_{GB}}^{(4)}
(x)+R^{(4)}(x)\Big[4(N+1)A^{\p\p}(z)\nn \\
&{}& -\, 2N(N+1){A^\p}^2(z)\Big]e^{2A(z)} +\cdots\,,
\end{eqnarray}
where $N=D-4$. In the following we restrict our
attention to $D=5$. The four-dimensional Planck scale is therefore
~\cite{IPN01c} \bea \label{5dPlanck} M_{Pl}^2 &\simeq&
M_{(5)}^{3}\int_{-\,\infty}^{\infty}
dz\,e^{-3A(z)}\left[1+4\alpha\kappa_5\,e^{2A(z)}
\left(2A^{\p\p}-{A^\p}^2\right)\right]\nn \\
&=& M_{(5)}^{3}\int_{-\,\infty}^{\infty}
dz\,e^{-3A(z)}\left(1+4\alpha\kappa_5\,e^{2A(z)}{A^\p}^2\right)
+M_{(5)}^3\,8\alpha\kappa_5\left[e^{-A(z)}{A^\p}\right]_{-\,\infty}^{+\,\infty}\,.
\eea It is obvious that $M_{Pl}^2$ is finite for
$A(z)=\ln\left(1+|z|/l\right)$. From Eq.~(\ref{fullRmumu}) one has
\be \label{fullRmumu2}
4\,\left(3{A^\p}^2-A^{\p\p}\right)=-\frac{8\,\kappa_5}
{3}\,e^{-2A(z)}\,V(\varphi) -8\varepsilon\, \left(A^{\p\p}-
{A^\p}^2\right)e^{2A(z)}{A^\p}^2\,. \ee Thus, with $\alpha>0$, the
inequality~(\ref{r-mumu}) may hold even in reverse order without
violating the positive energy condition, and the warp factor
$e^{-\,A(z)}$ need not be a constant. In particular, when the weak
energy condition $A^{\p\p}\geq - {A^\p}^2$ saturates, we find
\begin{equation}
16\,e^{2A(z)}{A^\p}^2\left(1-\varepsilon\,
e^{2A(z)}{A^\p}^2\right)= -\, \frac{8\,\kappa_5}{3}\,V(\varphi)\,.
\end{equation}
This can be satisfied when $V(\varphi)<0$, ${A^\p}^2>0$, and
$\varepsilon e^{2A(z)}{A^\p}^2 <1$. We have shown that in the
presence of a GB term there are some improvements over the no-go
theorem.

\subsection{Solutions with a bulk scalar}

For a non-constant scalar field, the bulk equations of motion
following from~(\ref{nogo1}) simplify to
\begin{eqnarray}
\kappa_{d+1}\,{\varphi^\p}^2 &=& 2
(d-1)\left({A^\p}^2+A^{\p\p}\right)\left(1-2\varepsilon {A^\p}^2
e^{2A(z)}\right)
\label{varphisq}\\
2\kappa_{d+1}\,V(\varphi)&=&-\,e^{2A(z)}(d-1)
\left[(d-1){A^\p}^2-A^{\p\p}\right]\nn \\
&{}& +\, (d-1)\varepsilon {A^\p}^2 e^{4A(z)}
\left[(d-2)\,{A^\prime}^2-2 A^{\p\p}\right]\,, \label{LambdaD}
\eea where $\varepsilon=(d-2)(d-3)\alpha\kappa_{d+1}$. We may
express these equations in the $y$ coordinate such that
$dy=e^{-\,A(z)} dz$. For $d=4$, and $\varphi=\varphi(y)$, $A'(z)
e^{A(z)}=A^\p(y)\equiv W(\varphi)$, $\partial W(\varphi)/\partial
\varphi\equiv W_\varphi$, Eqs.~(\ref{varphisq}), (\ref{LambdaD})
take the following form: \begin{eqnarray}
V(\varphi)&=&
\left(\frac{3\,W_\varphi^2}{2\kappa_5}-\Lambda_0\right)
\left(1-2\varepsilon W^2(\varphi) \right)^2 +\Lambda_0\,, \quad
\Lambda_0\equiv - \frac{3}{4\alpha}\,\frac{1}{\kappa_{5}^2}\,,
\label{potentialphi}\\
\varphi^\p(y)&=&\frac{6\,W_\varphi}{\kappa_5}\,
\left(1-2\varepsilon W^2(\varphi)\right)\,.
\end{eqnarray} The
above (super) potential~(\ref{potentialphi}), named due to its
resemblance with supergravity solution, for the $\varepsilon=0$
case, was analyzed in Ref.~\cite{Gubser99a}. In the following we
shall be interested in analyzing bulk solutions with
$\varepsilon>0$, without retaining explicit form of $A(y)$ (see
Refs.~\cite{Giovannini,kaku00b} for some relevant discussions).

Some non-singular solutions are found in Ref.~\cite{Rizos02a} by
satisfying $-22.2\lesssim \alpha\Lambda_0< -5/12$, in the units
$\kappa_5=1$. So in the bulk it might be possible to take
$V(\varphi)=\Lambda_0$ and $\left(1-2\varepsilon
W^2(\varphi)\right)=0$. In this limit, a domain wall solution
smoothly interpolates between two anti-de Sitter minima
$\varphi_{\pm}$ of the potential $V(\varphi)$ as $y\to \pm\,
\infty$. Because $\varphi^\p(y)$ vanishes in the bulk,
$V(\varphi)$ takes a bare value $V(\varphi_\pm)=\Lambda_0$. The
scalar field and warp factor simply become $ \varphi=\varphi_0$
and $A(y)=\pm\,\sqrt{-\,\Lambda_0}\,|y|$. One notes that
$1-2\varepsilon W_\varphi^2 \neq 0$ on the brane, rather this
becomes $1-2\varepsilon W_\varphi^2 /3>0$ due to an essential
$\delta$-function regularization. There is no any bulk singularity
for the (super) potential of the linear form
$W(\varphi)=\lambda_1\,\varphi+\lambda_2$~\cite{Zee}.

It might be desirable to know what would happen if $2\varepsilon
W^2(\varepsilon)\neq 1$ in the bulk$?$ Then, since \be
A^{\p\p}(y)=\varphi^\p(y)\,\frac{\partial W(\varphi)}{\partial
\varphi}= \frac{6\,W_\varphi^2}{\kappa_5}\,\left(1-2\varepsilon
W^2(\varphi)\right)\,, \ee the $A^{\p\p}(y)\geq 0$ condition holds
only if $2\varepsilon W^2(\varphi)<1$. In this case, there is no
improvement over the c--theorem. As we expect that $A^{\p\p}(y\neq
0)=0$, $W_\varphi$ should vanish in the bulk, and hence
$W(\varphi)$ becomes a constant function of scalar field, which
implies that a bulk scalar is not dynamical.


In the $\varepsilon=0$ case, the condition \be A^{\p\p}(y)=
\frac{6\,W_\varphi^2}{\kappa_5}\geq 0 \ee is used to prove a
c--theorem in Ref.~\cite{Gubser99a}. With $\varepsilon=0$, the
metric solution reads $A(y)\sim |y|/\ell$, where $\ell=\sqrt{6/(-
\Lambda)}$. In general, a domain-wall solution should interpolate
between $y\to -\infty$ (infrared region) where $A(y)$ is linear,
and $y\to \infty$ (ultraviolet region or AdS horizon) where $A(y)$
is again linear. But the condition $A^{\p\p}(y)\geq 0$ rules out
here the second possibility. So it is not possible to have
$A^\p(y)$ approach a positive constant as $y\to +\infty$ and a
negative constant as $y\to -\infty$. This monotonicity of
$A^\p(y)$ has been known as brane-world
``c--theorem''~\cite{Freedman99a,Gubser99a}. However, with
$\varepsilon>0$, and $2\varepsilon {A^\p}^2(y)=1$ as a bulk
solution (c.f.~[\ref{constphi2}]), warp factor gets the both signs
\be A(y)=\pm \frac{|y|}{l} \,. \ee Then it might be possible that
$A^\p(y)$ approaches a positive (negative) constant as $y\to
+\infty~(-\infty)$. Therefore, the brane-world ``c--theorem''
of~\cite{Freedman99a} may not be available to the $\alpha>0$ case.

\section{Discussion and Outlook}

If the Randall-Sundrum models are to be the low energy limits of
some fundamental (string) theory or yet-unknown theory of quantum
gravity, it is likely that the gravity equations include higher
curvature corrections such as a Gauss-Bonnet invariant. For a
background of branes coupled to matter sources and a GB
self-interaction term, we have presented some useful expressions
for the Neumann propagator in arbitrary $D$ dimensions and
analyzed the structure of graviton interaction. It is shown that
the RS model with a GB term in the bulk gives a massless graviton
on the brane as in the standard RS model. Perhaps this is one of
the most striking results of this paper. We have shown that for a
small GB coupling $\alpha$, so $\gamma< 1$, the brane-world
solutions are qualitatively similar to the RS solutions. We also
pointed out a possibility that the Newton's law is exact, other
than that such a behavior one would expect at large distance along
the brane, if one is allowed to take $\gamma$ in the order of
unity.

We have examined the general properties of the RS solutions
coupled to a bulk scalar and a GB term and have found that
fine-tuning is a generic feature of RS models. We analyzed the
energy conditions with warped space-time metrics in the bulk and
found that energy conditions are not violated by a GB term.
Meanwhile, we observed a new branch of solution with finite Planck
scale and no naked bulk singularity. More precisely, a bulk
singularity for the $\varepsilon=0$ solution, which may complement
the Hawking-Penrose singularity theorem~\cite{Hawking70a}, is
pushed for $\varepsilon>0$ to the singularity at the anti-de
Sitter horizon $z\to \pm\infty$, so is harmless as it might have
field theory interpretation. The new solutions with
$\varepsilon>0$ are also found useful to avoid some of the
previously known no--go arguments for the RS brane-world
compactifications.

In this paper, we have performed the calculations with an
assumption that the scalar modes of the metric fluctuations
decouple from the tensor modes. This is a possible one, because,
at least for the flat RS branes, there are no delta function
sources for the scalar and vector modes~\cite{IPN01c}. In the
presence of several co-dimension one branes, such a decoupling
certainly requires a moduli stabilization and it is possible that
the physics responsible for this stabilization would modify the
analysis for the tensor structure of the graviton propagator,
which is not analyzed here due to this subtlety. This treatment
might require more general assumptions that there is a bulk scalar
coupled to brane gravity, and the branes are (anti-)de Sitter. We
hope to return to this point in future publication.

\section*{Acknowledgements}

It is a great pleasure to thank M. Blau, G. Gabadadze and Carlos
Nu\~nez for fruitful discussions and comments on the subject. This
work was supported in part by the BK21 program of Ministry of
Education. One of us (IPN) also acknowledges partial support from
the SeoAm Foundation and wishes to thank the Abdus Salam ICTP for
a kind hospitality where part of the work was done.

\section*{Appendix A: Metric variations with quadratic curvature terms}

\renewcommand{\theequation}{A.\arabic{equation}}
\setcounter{equation}{0}

The starting brane-world action is the Lagrangian of gravity which
has a general form in $D$ dimensions, including quadratic-order
curvatures,
\begin{eqnarray}
S&=&\int_{M} d^{D}
x\sqrt{-g_{D}}\left[\frac{R}{\kappa_{D}}-2\Lambda +\left(\alpha
R^2+\beta R_{ab}R^{ab} +\gamma
R_{abcd}R^{abcd}\right)\right]\nn\\
&{}&+\int_{\partial M}^{\mbox{i'th~brane}} d^{D-1} x
\sqrt{-g_{(D-1)}}\, \big({\cal L}^{bdry}_m - \Lambda_i(z)\big)\nn
\\
&{}& +\, \int d^d x \sqrt{-g_{d}}\, (-\T) \,, \label{abceffective}
\end{eqnarray}
where $a,b,\cdots$ denote the $D$-dimensional space-time indices.
For $D=4$, the action~(\ref{abceffective}) will be free of massive
spin-$2$ ghost only if $\beta+4\gamma=0$~\cite{IPN01d}. For $D>4$,
however, one requires that $\alpha=-\beta/4=\gamma$, which is the
Gauss-Bonnet relation. The graviton equations derived by varying
the above action with respect to $g^{ab}$ may be expressed in the
following form
\begin{eqnarray}
&&\sqrt{-g_{D}}\left(\kappa_{D}^{-1}\,G_{ab}+ H_{ab}+\Lambda
g_{ab}\right)\nn \\
&&~~~~~= -\frac{1}{2}\,\sum_{i=1}^{D-4} \Lambda_i\,
\sqrt{-g_{D-1}^{(z_i=0)}}\, \delta(z_i)\,
\delta_a^p\delta_b^q\, g_{pq}^{(z_i=0)}\nn \\
&&~~~~~~~~~ -\,\frac{\T}{2}\,\sqrt{-
g_{d}\,^{(z_1,z_2,\cdots,z_n=0)}} \,\delta(z_1)\delta(z_2)\cdots
\delta(z_n)\delta_a^\mu\,\delta_b^\nu\,g_{\mu\nu}^{z_1,z_2,\cdots,z_n=0}\,,
\end{eqnarray}
where $H_{ab}$
\begin{eqnarray}
H_{ab} &=&-\frac{1}{2}\,g_{ab} (\alpha R^2+\beta R_{cd}R^{cd}+
\gamma R_{cdef}R^{cdef})\nn\\
& &+2\big[ \alpha R R_{ab}+\beta R_{acbd}R^{cd} +
 \gamma (R_{acde}R_b\,^{cde}-2R_a\,^c R_{bc}+2R_{acbd}R^{cd})\big]\nn\\
& &-(2\alpha+\beta+2\gamma)(\nabla_a\nabla_b R - g_{ab}\nabla^2R)+
(\beta+4\gamma)\nabla^2\left(R_{ab}-\frac{1}{2}\,g_{ab}
R\right)\,.
\end{eqnarray}
The curvature derivatives vanish for a Gauss-Bonnet invariant
${\cal R}_{GB}^2$ (i. e., with $4\alpha=-\beta=4\gamma$). Then the
tensor $H_{ab}$ reduces to the second order Lovelock tensor
\begin{eqnarray}\label{Habijk}
H_{ab}&=&-\frac{\alpha}{2}\, g_{ab}{\cal R}_{GB}^2 + 2\alpha
\left(R R_{ab}- 2 R_{acbd}R^{cd}
+ R_{acde}R_b\,^{cde}-2R_a\,^c R_{bc}\right)\nn \\
&=& 2\alpha\left( I_{ab}- 4 J_{ab}+ K_{ab}\right)\,,
\end{eqnarray}
where the quantities $I_{ab}, ~ J_{ab},~ K_{ab}$ are defined
below. The linearized form of the curvatures are

(i) Variation of $G_{ab}$ \be \delta G_{ab}=\delta
R_{ab}-\frac{1}{2}\, \bar{g}_{ab}\,\delta R -\half \,h_{ab}\R \,.
\ee

(ii) Variation of $I_{ab}\left(\equiv R\left(R_{ab}-\frac{1}{4}\,
g_{ab}R\right)\right)$ \be \delta I_{ab}= G_{ab}\,\delta R+\R
\,\delta R_{ab}-\frac{1}{4}\, h_{ab} \R^2\,. \ee

(iii) Variation of $J_{ab}\left(\equiv R_{acbd} R^{bd}
-\frac{1}{4}\,g_{ab} R_{cd} R^{cd}\right)$ \bea \delta J_{ab}&=&
\delta R_{apb}\,^q \R^p\,_q+\R_{apbq} \delta R^{pq}
-\half \bar{g}_{ab}\R^{pq}\delta R_{pq}\nn \\
&{}&-\Big(\R_{apb}\,^r \R_{qr}-\half \bar{g}_{ab}
\R_{pr}\R_q\,^r\big)h^{pq} -\frac{1}{4}\, h_{ab}\R_{pq}\R^{pq}\,.
\eea

(iv) Variation of $K_{ab} \left(\equiv R_{acde} R_b\,^{cde}
-\frac{1}{4}\,g_{ab} R_{cdef} R^{cdef} -2 R_a\,^c R_{bc}
+2R_{acbd} R^{cd}\right)$
\begin{eqnarray}
\delta K_{ab}&=& 2\R_{(a}\,^{pq}{}_{|r|}\,\delta
R_{b)pq}\,^r-\half\, \bar{g}_{ab} \R^{pqr}\,_s \delta
R_{pqr}\,^s\nn \\
&{}&+\, 2\R^c\,_d\delta R_{acb}\,^d + 2\R_{acbd}\delta
R^{cd}-4\R_{(a}\,^c \delta R_{b)c}\nn \\
&{}&-\, \Big(\R_{acpq} \R_{bd}\,^{pq} -\half\,
\bar{g}_{ab}\R_{cpqr}\R_d\,^{pqr}+2\R_{acb}\,^e \R_{de}
-2\R_{ac}\R_{bd}\Big)h^{cd}\nn \\
&{}&-\, \frac{1}{4}\,h_{ab}\R_{pqrs}\R^{pqrs}\,. \end{eqnarray}
The quantities defined with bar are to be taken about their
background values. In the following, we adopt a slightly different
scheme of linearization, which we find more convenient to use for
a warped bulk geometry. The linearized equations take the form
\begin{equation}
\label{newlinearized} \delta{\hat G}_{ab}+\kappa_D\,\delta{\hat
H}_{ab}=0\,,
\end{equation}
where
\begin{eqnarray}
\delta{\hat G}_{ab}&=&\delta G_{ab}-\bar{G}_{ac}h^c\,_b =\delta
G_{ab}-\kappa_{D}\delta T_{ab} =\delta R_{ab}-\frac{1}{2}\,
\bar{g}_{ab}\delta R-\R_{ac}h^c\,_b \\
\delta{\hat I}_{ab}&=&\delta I_{ab}-\bar{I}_{ac} h^c\,_b
=\bar{G}_{ab}\delta R+\R \delta R_{ab}-\R \R_{ac} h^c\,_b \\
\delta{\hat J}_{ab}&=&\delta J_{ab}-\bar{J}_{ac}h^c\,_b = \delta
R_{apb}\,^q \R^p\,_q+\R_{apbq} \delta R^{pq}
-\half \bar{g}_{ab}\R^{pq}\delta R_{pq}\nn \\
&{}&-\Big(\R_{apb}\,^r \R_{q r}-\half \bar{g}_{ab}
\R_{pr}\R_q\,^r\big)h^{pq}
-\R_{apcq}\R^{pq} h_b\,^c \\
\delta{\hat K}_{ab}&=& \delta K_{ab}-\bar{K}_{ac} h^c\,_b\nn \\
&=&2\R_{(a}\,^{pq}{}_{|r|}\,\delta R_{b)pq}\,^r-\half \bar{g}_{ab}
\R^{pqr}\,_s \delta R_{pqr}\,^s+2\R^c\,_d\delta R_{acb}\,^d\nn \\
&{}&+\, 2\R_{acbd}\delta
R^{cd}-4\R_{(a}\,^c \delta R_{b)c}\nn \\
&{}&-\,\Big(\R_{acpq} \R_{bd}\,^{pq} -\half\,
\bar{g}_{ab}\R_{cpqr}\R_d\,^{pqr}+2\R_{acb}\,^e \R_{de}
-2\R_{ac}\R_{bd}\Big)h^{cd}\nn \\
&{}&-\,\Big(\R_{apqr}\R_c\,^{pqr}+2\R_{apcq}\R^{pq}-2\R_a\,^p
\R_{cp}\Big) h_b\,^c
\end{eqnarray} In
using~(\ref{newlinearized}), the RS fine-tuned relations do not
appear explicitly in the equations of motion.

\section*{Appendix B: Useful Identities with Warped Space-Time Metrics}
\renewcommand{\theequation}{B.\arabic{equation}}
\setcounter{equation}{0} Let us consider the function defined by
$A(z)=\log (k|z|+1)$, where $k$ in the inverse $AdS$ curvature
scale, $k\equiv 1/L$. Since $A(z)$ is a function of $|z|$, we have
$\partial_zA=A'\,\partial_z|z|$, where
$\partial_z|z|=2\Theta(z)-1$, $A'$ is derivative of $A$ with
respect to its argument $|z|$, $\Theta(z)$ is the Heaviside
function. In $D$ space-time dimensions, the warp factor $A(z)$, as
a general solution of the Einstein field equations modified by a
Gauss-Bonnet term, reads \be A(z)=\log\left(\sum_{i=1}^{D-4}
k_i|z_i|+1\right) \,. \ee For simplicity we assume that
$k_1=k_2=\cdots=k$. A straight forward simplifications would rise
to give the following results \bea
{A'}^2&=&e^{-2A(z)}\,k^2\,\sum_{i}
\left(\partial_{z_i}|z|\right)^2 =e^{-2A(z)}\,k^2\,(D-4)
\,.\label{Aprime}\\
A^{\p\p}&=& -e^{-2A(z)}\,k^2\,(D-4)+e^{-A(z)}\,2 k
\sum_{i}\delta(z_i)\,. \label{Adoubleprime} \eea Some useful
identities that hold among the cross terms are \bea
\label{identities}
\partial_{z_j}\partial_{z_k} A\cdot \partial^{z_j} A\,\partial^{z_k} A
&=&e^{-3A(z)}\,2k^3\,\sum_{i}
\delta(z_i)- e^{-4A(z)}\,k^4\,(D-4)^2\,.\\
\partial_{z_j}\partial_{z_k}A\,\partial^{z_j}\partial^{z_k}A\,
&=& - e^{-2A(z)}\,4k^2\sum_{i\neq j}\delta(z_i)\,\delta(z_j)
+\left(A^{\p\p}\right)^2\,.\\
\partial_{z_j}\partial_{z_k}A+\partial_{z_j}A\partial_{z_k}A&=&e^{-A}\,k
\sum_i\left(\partial_{z_j}\partial_{z_k}|z_i|\right)
=e^{-A(z)}\,2k\,\delta(z_i)\,\delta_{z_j z_k}\,.\\
\partial_{z_k} A\,\partial^{z_k} h_{\mu\nu}&=&e^{-A(z)}\,k\sum_{i} sgn(z_i)\,
\partial_{z_i} h_{\mu\nu}\,.\\
\sum_i\left(\partial_{z_j}\partial_{z_k}|z_i|\right)\partial^{z_j}
\partial^{z_k} h_{\mu\nu}&=&2\sum_i \delta(z_i)\,\partial_{z_i}^2
h_{\mu\nu}\,. \eea

\section*{Appendix C: Linear Expansions with the Gauss-Bonnet Term}
\renewcommand{\theequation}{C.\arabic{equation}}
\setcounter{equation}{0}

In the gauge $h_\mu^\mu=0=\partial_\nu h^{\mu\nu}$, the first
order linearized equations for $h_{\mu\nu}$ take the general form
\begin{eqnarray}\label{labelc1}
\delta \hat{G}_{\mu\nu}&=&\delta G_{\mu\nu} -\delta
{T^{(0)}}_{\mu\lambda} h_\nu^\lambda =\bigg(-\half\,
\partial_\lambda^2 -\half\,\partial_{z_i}^2 + \frac{D-2}{2}\,
\partial^{z_i}A\, \partial_{z_i} \bigg) h_{\mu\nu}\,,\nn\\
\delta \hat{H}_{\mu\nu}&=& 2(D-4)\alpha
e^{2A}\bigg[\bigg(\frac{(D-5)}{2}\,
\partial_{z_i} A\,\partial^{z_i} A - \partial_{z_i}\partial^{z_i} A\bigg)
\left(\partial_\lambda^2+\partial_{z_i}^2\right)\nn\\
&&+\partial^{z_k} A \bigg((D-3)\partial_{z_i}\partial^{z_i} A -
\frac{(D-3)(D-4)}{2}\,
\partial_{z_i} A\, \partial^{z_i} A\bigg) \partial_{z_k}\nn\\
&& +\big(\partial_{z_i}\partial_{z_j} A + \partial_{z_i} A\,
\partial_{z_j} A\big)
\partial^{z_i}\partial^{z_j}\bigg]h_{\mu\nu}\,,
\end{eqnarray}
where as defined above $i,j= 1,2,\cdots,(D-4)$ count the number of
extra (transverse) coordinates. In using the metric solution
$A(z)=\log\left(\sum_{i=1}^{D-4}k|z_i|+1\right)$, the linearized
fluctuations $h_{\mu\nu}$ simplify to \bea\label{main-linear}
&&\frac{1}{\kappa_{D}}\Big[-\Box_4-\Box_z+(D-2)k
e^{-A(z)}\sum_i^{D-4} sgn(z_i)
\partial_{z_i}\Big]h_{\mu\nu}\nn \\
&&+2\alpha\, (D-4)\bigg[(D-3)(D-4)k^2\left(\Box_4+\Box_z\right)-4k
e^{A(z)}\sum_{i=1}^{D-4}
\delta(z_i)\Box_4 -4k e^{A(z)}\sum_{i\neq j}^{D-4}\delta(z_i)\partial_{z_j}^2\nn \\
&&+(D-3)k^2\Big(4\sum_{i=1}^{D-4}\delta(z_i)-(D-2)(D-4)k
e^{-A(z)}\Big)\sum_{i=1}^{D-4} sgn(z_j)
\partial_{z_j}\bigg] h_{\mu\nu}= 2 T_{\mu\nu}^{(m)}\,,
\eea where $\Box_z=\partial_{z_1}^2+\partial_{z_2}^2+\cdots
+\partial_{z_{D-4}}^2$. By redefining metric fluctuations as
$h_{\mu\nu}=e^{(D-2)A(z)/2}\, \tilde{h}_{\mu\nu}$, one can remove
from the first square bracket in Eq.~(\ref{main-linear}) the
single (linear) derivative term, and the kinetic term will have a
canonical form. After this rescaling, linearized equations take
the form \bea\label{linear-in-D}
&&\frac{1}{\kappa_{D}}\left[-\Box_4-\Box_z-(D-2)k e^{-A(z)}\sum
\delta(z_i)
+\frac{(D-4)(D-2)D\,k^2}{4}\, e^{-2A(z)}\right]\tilde{h}_{\mu\nu}\nn \\
&&+(D-4)\,\alpha e^{2A(z)}\bigg[2(D-3)(D-4)k^2 e^{-2A(z)}
\left(\Box_4+\Box_z\right)
-8k e^{-A(z)}\sum_i\delta(z_i)\,\Box_4 \nn \\
&&-8k e^{-A(z)}\sum_{i\neq j}\delta(z_i)\partial_{z_j}^2
+8\,k^2e^{-2A(z)}\Big((D-3)\sum_i
sgn(z_i)\delta(z_i)\partial_{z_i}
-\sum_{i\neq j}\delta(z_j)sgn(z_i)\partial_{z_i}\Big)\nn \\
&&-4(D-2)\,k^2 e^{-2A(z)} \bigg(\frac{D(D-3)(D-4)^2}{8}\,
k^2e^{-2A(z)}
+2\sum_{i\neq j}\delta(z_i)\,\delta(z_j)\nn \\
&&~~~~~~~~~~~~~
-\left(D^2-8D+18\right)\,e^{-A(z)}\sum_i\delta(z_i)\bigg)
\bigg]\tilde{h}_{\mu\nu}=2\,T_{\mu\nu}^{(m)}\,. \eea We have
implemented these expressions in the bulk part of the text.

\section*{Appendix D: Green Function in $D$ Dimensions}
\renewcommand{\theequation}{D.\arabic{equation}}
\setcounter{equation}{0}

In Sec. 3, we made use of several analytic properties of bessel
functions. Here we summarize some of the technical derivations
that were involved. By defining in Eq.~(\ref{propa1}) $\G_p = (z
z'/ L^2)^{(D-1)/2}\, \hat{\G}_p$ and $p^2 = -\,q^2$, we arrive to
\begin{eqnarray}
\left(1-\gamma\right)\left(z^2\partial^2_z+z\partial_z+q^2 z^2
-\frac{(D-1)^2}{4}\right) \hat{\G}_p(z,z^\p) =L\,
z\,\delta(z-z^\p) \label{greeneq2}\,.
\end{eqnarray}
One can also derive the $D$ dimensional boundary condition at
$z=L$, analogous to the Neumann condition on the gravitational
field that $\partial_z \G_D(x, x')|_{z=L}=0$, to be
\begin{equation}
\bigg(z\partial_{z} +\frac{D-1}{2}+\frac{2}{(D-3)}
\frac{\gamma}{(1-\gamma)}\, q^2
z^2\bigg)\hat{\G}_p(z,z^\p)|_{z=L}=0\,. \label{greenbc}
\end{equation}
This is a boundary condition implied by continuity of the
derivative fields on the brane. Equation~(\ref{greeneq2}) further
implies the following matching conditions at $z=z'$:
\begin{equation}
\hat{\G}_<|_{z=z^\p}=\hat{\G}_>|_{z=z^\p}\,,\quad
\partial_z(\hat{\G}_>-\hat{\G}_<)|_{z=z^\p}= \big(1-\gamma\big)^{-1}
\frac{L}{z^\p} \label{match1}\,.
\end{equation}
One can find the general solution of~(\ref{greeneq2}) by
satisfying~(\ref{match1}) and~(\ref{greenbc}). The most general
solution of~(\ref{greeneq2}), for $z < z'$ is
\begin{equation}
\hat{\G}_{z<z'} = iA(z^\prime)\Big[\big(J_{\nu-1}+ \chi\,(q L)
J_{\nu}\big)H^{(1)}_{\nu}(qz)-\big(H^{(1)}_{\nu-1}(q L)+\chi\, (q
L)\, H^{(1)}_{\nu}\big)J_{\nu}(qz)\Big]\,, \label{greenleft}
\end{equation}
where $\chi = \gamma/{((\nu-1)\,(1-\gamma))}$, $\nu=(D-1)/2$, and
$H^{(1)}_{\nu}(q L)=J_{\nu}(q L)+iY_{\nu}(q L)$ is the Hankel
function of the first kind. For $z>z'$, one can use the following
boundary condition at the AdS horizon, $z=\infty$:
\begin{equation}
{\hat G}_{z>z'}\,=\,B(z')\,H^{(1)}_{\nu}(qz) \label{Hawkingbc}\,.
\end{equation}
Following~\cite{Giddings00a,IPN01b}, we solve
Eqs.~(\ref{match1},\ref{greenleft},\ref{Hawkingbc})) and arrive to
a $D$-dimensional Neumann propagator
\begin{eqnarray}
\G_{D}(x,z;x,z^\p)&=& -\, (1-\gamma)^{-1} \frac{i\pi}{2 L^3}(z
z^\p)^{\nu} \int\frac{d^{2\nu}p}{(2\pi)^{2\nu}} e^{ip(x-x')}\,
\bigg[J_{\nu}(qz_<)H^{(1)}_{\nu}(qz_>)\nn\\
&{}&-\,\left(\frac{J_{\nu-1}(q L) +\chi\,(q L)\, J_{\nu}(q
L)}{H^{(1)}_{\nu-1}(q L) +\chi\,q L\, H^{(1)}_{\nu}(qL)} \right)
H^{(1)}_{\nu}(qz) H^{(1)}_{\nu}(qz') \bigg]\,. \label{greenpro3}
\end{eqnarray}
This gives the results in Ref.~\cite{IPN01b} for $D=5$, and that
of Ref.~\cite{Giddings00a} with $\gamma=0$ and $D=d+1$. When an
argument of the propagator is at $z'= L $, ~(\ref{greenpro3}) can
be reduced to Eq.~(\ref{greengen}) given in the text. When both
arguments are on the brane, Eq.(\ref{greengen}), after Bessel
expansions and some algebraic manipulations, would simplify to
yield
\bea &&{\cal G}_D(x,L;x',L)\simeq (1-\gamma)^{-1}\int\frac{d^{D-1}
p}{(2\pi)^{D-1}}\, e^{ip(x-x')}\,\frac{1}{q}\,
\bigg[\frac{1-\gamma}{1+\gamma}\, \frac{(D-3)}{qL}
+\frac{\left[(D-4)+(D-6)\gamma\right]}
{2(D-5)}\nn \\
&&~~~~\times\frac{(1-\gamma)}{(1+\gamma)^2}\,qL +{\cal
O}\big((qL)^3\big)+\cdots+\frac{(1-\gamma)^2}
{(1+\gamma)^2}\,\frac{{(qL)}^{D-4}}{C_1}\,
\ln\Big(\frac{qL}{2}\Big)\bigg]\,, \label{propagator2} \eea where
$C_1$ is a dimensional constant ($C_1=1, 4, 64,\cdots $, for $D=5,
7, 9,\cdots$). The the last expression involving logarithmic term
would be absent when the number of total dimensions $D$ is even.



\begin{thebibliography}{100}

\bibitem{Randall1} L. Randall and R. Sundrum, Phys. Rev. Lett. {\bf 83}
(1999) 3370, arXiv:hep-ph/9905221.

\bibitem{Randall2} L. Randall and R. Sundrum, Phys. Rev. Lett. {\bf 83}
(1999) 4690, arXiv:hep-th/9906064.

\bibitem{ArkaniHamed99b} N. Arkani-Hamed, S. Dimopoulos, G. Dvali and N.
Kaloper, Phys. Rev. Lett. {\bf 84} (2000) 586,
arXiv:hep-th/9907209.

\bibitem{Garriga99a} J. Garriga and T. Tanaka, Phys. Rev. Lett.
{\bf 84} (2000) 2778, arXiv:hep-th/9911055.

\bibitem{Kaloper99a} N.~Kaloper,
Phys. Lett. B {\bf 474} (2000) 269, arXiv:hep-th/9912125.

\bibitem{Csaki00a}  C.~Cs$\acute{a}$ki, J. Erlich, T.J. Hollowood and
Y. Shirman, Nucl. Phys. B {\bf 581} (2000) 309,
arXiv:hep-th/0001033.

\bibitem{Giddings00a} S.~B.~Giddings, E.~Katz and L.~Randall,
J. High Energy Physics {\bf 0003} (2000) 023,
arXiv:hep-th/0002091.

\bibitem{JEKim00b} J. E. Kim and H. M. Lee, Nucl. Phys. B {\bf 602}
(2001) 346, arXiv:hep-th/0010093; Erratum-ibid, Nucl. Phys. B {\bf
619} (2001) 763.

\bibitem{IPN01b} Y.~M.~Cho, I.~P.~Neupane, and P.~S.~Wesson,
Nucl. Phys. B {\bf 621} (2002) 388, arXiv:hep-th/0104227.

\bibitem{Deser85a} D.~G.~Boulware and S. Deser, Phys. Rev. Lett.
{\bf 55} (1985) 2656.

\bibitem{Metsaev87a} R.~R.~Metsaev and A.~A.~Tseytlin, Nucl. Phys. B
{\bf 293} (1987) 385;

D. J. Gross and J. H. Sloan, Nucl. Phys. B {\bf 291} (1987) 41.

\bibitem{Kashima} A. Lukas, B. A. Ovrut and D. Waldram,
{\tt Nucl. Phys.} B {\bf 532}, 43 (1998), arXiv:hep-th/9710208; K.
Kashima, {\em Prog. Theor. Phys.} {\bf 105}, 301 (2001),
arXiv:hep-th/0010286.

\bibitem{JEKim00a} J. E. Kim, B. Kyae and H. M. Lee, Nucl. Phys. B
{\bf 582} (2000) 296, arXiv:hep-th/0004005;

\bibitem{Zee} I. Low and A. Zee, Nucl. Phys. B {\bf 585} (2000) 395,
arXiv:hep-th/0004124.

\bibitem{Nojiri00c} S. Nojiri and S. D. Odintsov, Phys. Lett. B
{\bf 484} (2000) 119, arXiv:hep-th/0004097;
JHEP {\bf 0007} (2000) 049, arXiv:hep-th/0006232.

\bibitem{JEKim01a} J.~E.~Kim, B. Kyae and H.~M.~Lee, Phys.
Rev. D {\bf 64} (2001) 065011, arXiv:het-th/0104150.

\bibitem{IPN01c} I.~P.~Neupane, Class. Quant. Grav. 19 (2002) 5507,
arXiv:hep-th/0106100.

\bibitem{IPN01a} I.~P.~Neupane, Phy. Lett. B {\bf 512} (2001) 137,
arXiv:hep-th/0104226.

\bibitem{Mavro} N. E. Mavromatos and J. Rizos, Phys. Rev. D {\bf 62}
(2000) 124004, arXiv:hep-th/0008074.

\bibitem{IPN00} I. P. Neupane, {JHEP} {\bf 0009} (2000) 040,
arXiv:hep-th/0008190.

\bibitem{Deruelle} N. Deruelle and T. Dolezel,
Phys. Rev. D {\bf 62} (2000) 103502, arXiv:gr-qc/0004021.

\bibitem{Collins} H. Collins and B. Holdom, Phys. Rev. D {\bf 63}
(2001) 084020, arXiv:hep-th/0009127.

\bibitem{Giovannini} M. Giovannini, Phys. Rev. D {\bf 63} (2001)
064011, arXiv:hep-th/0011153; Phys. Rev. D {\bf 63} (2001) 085005,
arXiv:hep-th/0009172;
Phys. Rev. D {\bf 64} (2001) 124004, arXiv:hep-th/0107233.

\bibitem{Kaku00a} Z. Kakushadze, Phys. Lett. B {\bf 494} (2000) 302,
arXiv:hep-th/0009022;
O. Corradini, A. Iglesias, Z.~Kakushadze and P. Langfelder, Phys.
Lett. B {\bf 521} (2001) 96, arXiv:hep-th/0108055.

\bibitem{kaku00b} A. Iglesias and Z. Kakushadze, Int. J. Mod. Phys. A
{\bf 16} (2001) 3603, arXiv:hep-th/0011111.

\bibitem{0106203} K. A. Meissner and M. Olechowski, Phys. Rev.
Lett. {\bf 86} (2001) 3708, arXiv:hep-th/0009122; Phys. Rev. D
{\bf 65} (2002) 064017, arXiv:hep-th/0106203.

\bibitem{Nojiri00d} S. Nojiri and S. D. Odintsov, JHEP {\bf 0112}
(2001) 033, arXiv:hep-th/0107134;

B.~Abdesselam and N.~Mohammedi, Phys. Rev. D {\bf 65} (2002)
084018, arXiv:hep-th/0110143;

C. Charmousis and J.-F. Dufaux, Class. Quant. Grav. {\bf 19}
(2002) 4671, arXiv:hep-th/0202107.

\bibitem{Germani02a} C.~Germani and C.~F.~Sopuerta,
Phys. Rev. Lett. {\bf 88} (2002) 231101, arXiv:hep-th/0202060.

\bibitem{IPN02a} J. E. Lidsey, S. Nojiri and S. D. Odintsov, {JHEP},
{\bf 0206} (2002) 026, arXiv:hep-th/0202198;
Y. M. Cho and I.~P.~Neupane, Phys. Rev. D {\bf 66} (2002) 024044,
arXiv:hep-th/0202140; S. Nojiri, S. D. Odintsov and S. Ogushi,
Int. J. Mod. Phys. A {\bf 17} (2002) 4809, arXiv:hep-th/0205187.

\bibitem{CEGH} C.~Csaki, J.~Erlich, C.~Grojean and T.~Hollowood, Nucl.
Phys. B {\bf 584} (2000) 359, arXiv:hep-th/0004133;

\bibitem{Freedman99a} D.~Z.~Freedman, S. S. Gubser, K. Pilch and
N. P. Warner, Adv. Theor. Math. Phys. {\bf 3} (1999) 363,
arXiv:hep-th/9904017.

\bibitem{Malda00a} J. Maldacena and Carlos Nu\~nez,
Int. J. Mod. Phys. A {\bf 16} (2001) 822, arXiv:hep-th/0007018.

\bibitem{Rama02a} S. Dasgupta, R. Venkatachalapathy and S. K. Rama,
JHEP {\bf 0207} (2002) 061, arXiv:hep-th/0204136.

\bibitem{Gherghetta00a} T. Gherghetta and M. Shaposhnikov, Phys.
Rev. Lett. {\bf 85} (2000) 240, arXiv:hep-th/0004014.

\bibitem{Dvali98a} N. Arkani-Hamed, S. Dimopoulos and G.
Dvali, Phys. Lett. B {\bf 429} (1998) 263, arXiv:hep-ph/9803315.

\bibitem{Hawking70a} S. W. Hawking and R. Penrose, Proc. Roy.
Soc., London, A {\bf 314} (1970) 529.

\bibitem{Gibbons} G. W. Gibbons, ``Aspects of Supergravity Theories'',
GIFT Seminar 1984, pp. 123-146, (QCD161:G2:1984), {\tt
Print-85-0061} (CAMBRIDGE).

\bibitem{Gubser99a} O. DeWolfe, D. Z. Freedman, S. S. Gubser and
A. Karch, Phys. Rev. D {\bf 62} (2000) 046008,
arXiv:hep-th/9909134.

\bibitem{Rizos02a} N. E. Mavromatos and J. Rizos, Int. J. Mod. Phys. A
{\bf 18} (2003) 57, arXiv:hep-th/0205299; P. Bin\'etruy, C.
Charmousis, S. Davis and J.-F. Dufaux, Phys. Lett. B {\bf 544}
(2002) 183, arXiv:hep-th/0206089.

\bibitem{Sundrum00a} N. Arkani-Hamed, S. Dimopoulos, N. Kaloper
and R. Sundrum, Phys. Lett. B {\bf 480} (2000) 193,
arXiv:hep-th/0001197; S. Kachru, M. Schulz and E. Silverstein,
Phys. Rev. D {\bf 62} (2000) 045021, arXiv:hep-th/0001206; S. P.
de Alwis, Nucl. Phys. B {\bf 597} (2001) 263,
arXiv:hep-th/0002174.

\bibitem{IPN01d} I.~P.~Neupane, Class. Quant. Grav. {\bf 19} (2002) 1167,
arXiv:hep-th/0108194.


\end{thebibliography}
\end{document}